\documentclass[12pt]{iopart}
\newcommand{\beqn}{\begin{eqnarray}}
\newcommand{\eeqn}{\end{eqnarray}}
\def\half{\frac{1}{2}}

\def\ptl{\partial}
\newcommand{\beqar}{\begin{eqnarray}}
\newcommand{\eeqar}{\end{eqnarray}}
\newcommand{\llabel}[1]{\label{#1}}              

\newcommand{\labeq}[2]{ \begin{equation} \llabel{#1}{#2} \end{equation}}

\usepackage{graphicx}
\usepackage{epsf}
\usepackage{amssymb}
\usepackage{graphics,epsfig,placeins,subfigure,wrapfig}

\begin{document}
\title{Constraint propagation equations of the 3+1 decomposition of $f(R)$ gravity}
\author{Vasileios Paschalidis${}^1$, Seyyed~M.~H. Halataei${}^2$, \\ Stuart~L. Shapiro${}^{1,2,3}$ and Ignacy Sawicki${}^4$}
%

\address{
${}^1$ Department of Physics, University of Illinois at Urbana-Champaign,
Urbana, IL 61801, USA}  
\address{${}^2$ Department of Astronomy, University of Illinois at Urbana-Champaign,
Urbana, IL 61801, USA}
\address{${}^3$ NCSA, University of Illinois at Urbana-Champaign,
Urbana, IL 61801, USA}
\address{${}^4$ Institut f\"{u}r Theoretische Physik,Ruprecht-Karls-Universit\"{a}t Heidelberg, Philosophenweg 16, 69120 Heidelberg, Germany 
}
\vspace{0.7cm}
\address{Email: vpaschal@illinois.edu}

%
\date{\today}

\begin{abstract}

Theories of gravity other than general relativity (GR) can explain the observed cosmic
 acceleration without a cosmological constant.
One such class of theories of gravity is $f(R)$. Metric $f(R)$ theories have been proven to be
 equivalent to Brans-Dicke (BD) scalar-tensor gravity without a kinetic term ($\omega=0$). 
Using this equivalence and a 3+1 decomposition of the theory 
 it has been shown that metric $f(R)$ gravity admits a well-posed initial value problem. 
However, it has not been proven that the 3+1 evolution equations of metric 
$f(R)$ gravity preserve the (hamiltonian and momentum) constraints. In this paper we show that this
 is indeed the case. In addition, we show that the mathematical form of the constraint propagation equations
in BD-equilavent $f(R)$ gravity and in $f(R)$ gravity in both the Jordan and Einstein frames, is exactly the
same as in the standard ADM 3+1 decomposition of GR. 
Finally, we point out that current numerical relativity codes 
can incorporate the 3+1 evolution equations of metric $f(R)$ gravity by modifying the stress-energy tensor
 and adding an additional scalar field evolution equation. 
We hope that this work will serve as a starting point for relativists to develop fully dynamical codes for valid $f(R)$ models.


\end{abstract}


\maketitle
%
%

\section{Introduction}
\label{Introduction}

Since its formulation, Einstein's general theory of relativity (GR) has withstood extensive experimental and
observational scrutiny using tests that range from millimeter to solar system scales (see \cite{Will_review} 
and references therein). The discovery of the late-time cosmic 
acceleration 
\cite{SNcosmic_accel,Perlmutter} was a surprise, 
but one  which could be modelled within the minimally extended
 framework of $\Lambda$CDM 
\cite{Peebles,lrr-2001-1} -- GR with a positive cosmological constant. To 
this day this 
simple model remains in very good agreement with data from all competitive probes
 \cite{Komatsu:2010fb, Amanullah:2010vv, Reid:2009xm, Percival:2009xn}, which 
imply that approximately 70\% of the energy density of the universe
 is made up of a component which
 does not cluster and has an equation of state with
 pressure approximately equal to minus the energy density. 
While the simplest model for this component is indeed the cosmological
 constant, from the point of view of particle physics, 
its value implied by the measurements of the cosmological expansion
 is extremely low and requires a very high level of fine tuning.

A number of alternative models for dark energy have been proposed, 
most of which suffer from a similar fine-tuning problem to $\Lambda$CDM
 (see the review \cite{Copeland:2006wr}), but at least
 provide a set of alternatives against which to test the $\Lambda$CDM
 hypothesis. In this spirit, it is possible to imagine that, rather
 than proposing the existence of a new, exotic form of energy density, it
 is the theory of gravity which we use to interpret the cosmological
 data that must be modified.



 There are a number of proposed gravity theories which modify the
 dynamics at large distances, and  metric $f(R)$ theories 
of gravity (see, e.g., \cite{f_of_R_review,Sotiriou} and references
 therein) comprise one such class of modifications to GR. This class has attracted 
 considerable attention in recent years, perhaps due to the
simplicity of the modifications. Further motivation for the study of $f(R)$ gravity is reviewed
 in \cite{Sotiriou}; for other interesting alternatives,
see \cite{f_of_R_review} for Gauss-Bonet gravity, \cite{Mannheim2006340} for conformal gravity and
\cite{MaartensRev} for Brane-World gravity. 

The $f(R)$ formulation arises from a
 simple replacement of the Ricci scalar ($R$) in
the Einstein-Hilbert action, 
\labeq{EHaction}{
S = \frac{1}{16\pi}\int \sqrt{-g}d^4x (R-2\Lambda) + S_m(g_{\mu\nu},\psi_m),
}
where $g$ is the determinant of the metric tensor
 $g_{\mu\nu}$, $\Lambda$ the cosmological constant, $S_m$ the
 matter term in the action, 
and $\psi_m$ collectively denotes the matter fields,  
by an arbitrary function of the Ricci scalar, i.e., 
\labeq{fofRaction}{
S = \frac{1}{16\pi}\int \sqrt{-g}d^4x f(R) + S_m(g_{\mu\nu},\psi_m).
}
Note that throughout this work we adopt geometrized units, where $G=c=1$.

From Equations \eref{EHaction} and \eref{fofRaction} it is clear 
that GR is recovered for $f(R)=R-2\Lambda$. In metric $f(R)$ theories
 the connection symbols ${}^{(4)}\Gamma^i{}_{jk}$ are chosen to
 be the Christoffel symbols associated with the metric tensor, so that 
the action is a function of only the metric tensor and its derivatives. 
As a result, in metric $f(R)$ gravity only the metric tensor is truly dynamical. 
In Palatini $f(R)$ gravity the connections ${}^{(4)}\Gamma^i{}_{jk}$ are
 considered independent of the metric tensor, so that the action is
 a function of both the metric tensor and the connection symbols. Thus, 
in Palatini $f(R)$ both $g_{\mu\nu}$ and ${}^{(4)}\Gamma^i{}_{jk}$ are
 dynamical fields (see also \cite{Amendola:2010bk} for a new class of
 models which interpolate between the metric and Palatini formulations).
In this work we are concerned with metric $f(R)$ gravity only.

Early work on $f(R)$ theories \cite{Bergmann,Ruzmaikina,Buchdahl,Starobinsky1980}
 was mainly concerned with high-energy corrections to
 general relativity and their influence on the early universe 
(see in particular \cite{Starobinsky1980} where the first $f(R)$ model of inflation
 was proposed). The discovery of  cosmic acceleration 
\cite{SNcosmic_accel,Perlmutter} renewed the interest in $f(R)$ models, 
but now with modifications in the infra-red. A number of alternative models to
 GR have been proposed \cite{Capozziello0,Capozziello,Capozziello2,CDTT,Nojiri}. 
However, it was later shown that these models neither satisfy
 local gravity constraints \cite{Chiba,Olmo1,Olmo2} nor 
give rise to a standard matter-dominated era \cite{Amendola1,Amendola2}. 


General conditions for the cosmological viability of $f(R)$ models were derived in 
\cite{Amendola3} and it was later realized that the
 so-called Chameleon mechanism - the scalar degree of freedom becomes massive in dense environments and light in diffuse ones -
 can allow $f(R)$ gravity to satisfy Solar-System constraints \cite{Faulkner:2006ub,Hu_Iggy}. 
The key consequence of the Chameleon mechanism is that the modification to the metric inside galactic haloes
is suppressed: gravity returns to its general-relativistic behaviour. The functioning of the Chameleon mechanism has
 also been confirmed via N-body simulations of large-scale cosmological structure
 formation in \cite{Oyaizu:2008sr, Oyaizu:2008tb, Schmidt:2008tn,Ferraro:2010gh, Zhao:2010qy}, where it was shown that 
predictions for cluster abundance and the matter power spectrum return at small scales to those calculated within the $\Lambda$CDM framework.

A number of models that satisfy both Solar-System and cosmological constraints 
have been proposed in \cite{Faulkner:2006ub,Hu_Iggy,Li,Starobinsky2,Appleby,Deruelle,Cognola,Linder}, and it is now
known that for an $f(R)$ theory to be viable the following four constraints must be met \cite{f_of_R_review}:
\begin{enumerate}

\item $f,_{R} > 0 $ for $R \geq R_0$, where $R_0$ is the cosmological 
value of the Ricci scalar today. This condition is necessary 
for guaranteeing that the new scalar degree of freedom is not a ghost -- a field with negative kinetic energy.

\item $f(R) \rightarrow R$ for $R \gg R_0$. This condition
 is necessary for the presence of a matter-dominated era and to
 evade solar-system constraints. 

\item $f,_{RR} > 0 $ for $R \geq R_0$ in the presence of external matter. 
This condition ensures that the matter-dominated era is the stable
 solution for cosmology and that the solutions which satisfy solar
 system constraints are stable.

\item $0 < Rf,_{RR}/f,_R|_{r=-2}$, where $r = -Rf,_R/f$. 
This condition is necessary for the stability and presence of a late time de Sitter solution. 

\end{enumerate}

The existence of these requirements is a result
 of the fact that in $f(R)$ gravity the Ricci scalar is a full
 dynamical degree of freedom, which must behave in a manner similar to
 the Ricci scalar in GR, where it is controlled through a
 constraint ($R=-8\pi T$). These conditions ensure that in high-density
 environments, the so-called high-curvature solutions, 
 where $R\simeq 8\pi\rho$, are stable.


An additional constraint that any theory of gravity must satisfy is the existence of stable relativistic (neutron) stars. It was originally pointed
out in \cite{Frolov}, that many models of $f(R)$ theories reach a curvature singularity at a finite value of the 
scalar degree of freedom $f,_R$ which is not protected by the existence of a potential barrier. This value
 of the scalar field may be attained in the presence of relativistic matter. This same idea was used in \cite{Kobayashi} to argue 
that it is not possible to build spherically symmetric, i.e., non-rotating, relativistic stars in $f(R)$ theories of gravity. 
These works stimulated further interest and eventually numerical models of spherical relativistic stars in $f(R)$ gravity
 were explicitly constructed in \cite{Babichev,Upadhye_Hu,Babichev2}. There it
 was shown that building numerical models of neutron stars in $f(R)$ gravity is very sensitive to the treatment of boundary conditions. 

To our knowledge a stability analysis of non-rotating equilibrium models of neutron stars in the context of $f(R)$ theories has not been carried out yet. 
One may expect that the stability properties of relativistic stars in viable $f(R)$ gravity are the 
same as those in GR, because of condition 
3 above. However, given the subtleties that arise in obtaining relativistic stellar configurations in $f(R)$ theories 
due to the effective scalar degree of freedom it is natural to expect that the back-reaction of the scalar field
 will affect the stability, too. In addition, it would be interesting to explore
 the existence and stability of rotating neutron stars and how $f(R)$ gravity affects the criterion for the onset of the bar mode, r-mode and other
non-axisymmetric instabilities \cite{LRS93a,LRS94,ChandrasekharEllips,ShibataBarmode,SaijoBarmode,KokkotasRev}. 
Furthermore, it is intriguing to study gravitational radiation arising from compact stars, both in isolation and in binary systems. Included 
in this list are
neutron star -- neutron star \cite{DuezNSNSReview}, black hole--black hole \cite{2010CQGra..27k4004H}, black hole--neutron star 
\cite{Rantsiou08, Loffler06, Faber, Faber06, Shibata06, Shibata07, Shibata08,Yamamoto08,Etienne08a, Etienne08, 
Duez08,2009PhRvD..79d4030S,2009PhRvD..79l4018K,2010AAS...21530001M,2010arXiv1006.2839C,2010CQGra..27k4106D,
2010arXiv1007.4160P,2010arXiv1007.4203F,2010arXiv1008.1460K} and white-dwarf--neutron star binaries \cite{WDNS_PAPERI,WDNS_PAPERII}.

Some of these studies can be carried out analytically via perturbation theory, and some require direct numerical simulations. 
One of the main points we make in this work is that current numerical relativity techniques (see texts by Baumgarte and Shapiro
\cite{BSBook} and Alcubierre \cite{AlcuBook} and references therein), i.e., 
the solution of the Einstein equations by computational means, 
should be able to handle the equations of $f(R)$ gravity 
straightforwardly. In particular, the minimum requirement
is to modify the stress-energy tensor and add a new scalar field evolution equation. 
However, to achieve long-term stable numerical integration of any set of partial differential equations, 
well-posedness of the Cauchy (or initial value) problem must be guaranteed.

Unlike GR, the field equations of metric $f(R)$ gravity in the so-called Jordan frame are 4th order (see \Sref{fofRequations}). 
Nevertheless these theories can be cast in 2nd-order form, by promoting $f,_R$ (the derivative of $f(R)$ 
with respect to $R$) to an effective dynamical scalar degree of freedom. Alternatively, metric $f(R)$ gravity can be 
reduced to second-order form by a transformation of the $f(R)$ action to a Brans-Dicke (BD) \cite{BransDicke} action 
with $\omega=0$ \cite{Chiba}. This means that $f(R)$ gravity is equivalent to BD gravity without a kinetic term. 
Exploiting this equivalence and the 3+1 decomposition approach of \cite{Salgado06}, it was 
demonstrated in \cite{fofR_well_posed} that metric $f(R)$ gravity admits a well-posed initial value problem. As in 3+1 GR,
to solve the initial value problem, first one solves the 3+1 constraint equations 
to obtain initial data and then uses the 3+1 evolution equations to advance the initial data in time. For this approach to yield a consistent
solution of the covariant (4D) field equations, the 3+1 evolution equations must preserve the constraints of the theory. 
 To prove this one has to derive the evolution equations of the constraints, which are often referred to as the constraint propagation equations,
and show that if the constraint equations are initially satisfied, they must be satisfied for all times. 
To our knowledge this has never been demonstrated for a 3+1 formulation of $f(R)$ gravity and in this work we show that this is indeed the case. 



To date there are two methods for deriving 3+1 constraint propagation equations.
One approach is to take the time derivative of the constraint equations in 3+1 form and then replace the time derivatives
 of all dynamical variables by using the evolution equations for these variables. We call this the 3+1 or ``brute force'' method.
This is a rather tedious approach and to our knowledge, it has been performed in GR only for vacuum spacetimes in \cite{VasPas_formulations}.
A pedagogical example that explains the ``brute force'' method is given in section II of \cite{VasPas_formulations}, 
and more involved applications involving Maxwell's equations can be found in \cite{Calabrese}.

The other approach, which is  more elegant, takes advantage of the Bianchi identities. We call this the Frittelli method \cite{Frit97} (see also \cite{AlcuBook}). 

However, the equations derived in \cite{Frit97} were not cast in pure 3+1 
language. Here and throughout this paper by ``pure 3+1 language'' we mean that a given equation is written solely 
in terms of scalars and purely spatial objects and their derivatives. Yoneda and Shinkai \cite{YonShin2,2004GReGr..36.1931S} have derived 
the Arnowitt, Deser, Misner (ADM) constraint propagation equations in pure 3+1 language but they did not indicate how they arrived at their result.

In this work we employ the Frittelli approach to derive the constraint propagation equations of $f(R)$ gravity
and cast the resulting equations in pure 3+1 language. 
We show that the mathematical form of the constraint propagation equations 
is the same as that of the standard ADM formulation of GR. We also demonstrate that this result holds true
both in the Jordan and the Einstein frames of metric $f(R)$  gravity, as well as for the BD-equivalent version of metric $f(R)$ gravity. 
Finally, we compare our equations with published results of the constraint propagation equations 
derived using the 3+1 approach and show that the expressions obtained via both approaches agree. 

While none of our results are surprising they serve to prove that f(R) gravity is self-consistent. Moreover, it is revealing 
to demonstrate how previous results from GR can be extended to alternative theories of gravity and the consistency between 
alternative approaches. Finally, obtaining the extended constraint propagation equations in pure 3+1 form
may prove useful for performing 3+1 numerical simulations, where constraint preservation can be used as a check on the integration.

This paper is organized as follows. In \Sref{fofRequations} we review the field equations of generic metric $f(R)$ models.
 In \Sref{con_pres} we provide a simple pedagogical argument (see also \cite{WeinbergGR,BSBook}) to demonstrate the basic idea of
 constraint preservation in the context of GR. In \Sref{3p1_intro} we review the 3+1 decomposition of the BD-equivalent metric $f(R)$ equations.
In \Sref{General_Con_prop} we employ the Frittelli method and use the results of \Sref{3p1_intro} to derive the 3+1 metric $f(R)$
constraint propagation equations. In \Sref{3p1_lang_con_prog}
we cast our generalized evolution equations of the constraints in pure 3+1 language. In \Sref{jordan_Einstein} we argue
that the 3+1 constraint propagation equations of $f(R)$ gravity in both the Jordan and the Einstein frames can be cast in the same
form as that in the 3+1 BD-equivalent version of $f(R)$ theories. Finally, we summarize our work in \Sref{summary}.

\section{$f(R)$ field equations}
\label{fofRequations}

As in GR, the fundamental quantity in $f(R)$ gravity is the spacetime metric tensor $g_{\alpha\beta}$
\labeq{metric1}{
ds^2 = g_{\alpha\beta}dx^{\alpha}dx^{\beta},
}
where $ds$ is the line element, and $x^{\alpha}$ denote the spacetime coordinates. 
Here and throughout this paper
Greek indices run from $0$ to $3$, while Latin indices run from $1$ to $3$. 

The goal of the theory is to determine the metric given a mass-energy distribution. 
Because of the existence of an additional scalar degree of freedom in the gravitational field sector, 
it is possible to formulate the field equations 
of $f(R)$ theory in many ways, depending on the amount of mixing between these two fields. 
We will discuss three such formulations: the Jordan frame, 
the Einstein frame, and the BD-equivalent formulation. 


The Jordan frame and Einstein frame formulations have different metrics as dynamical variables.  
The two metric tensors are related via a conformal transformation

\labeq{conf_transf_gen}{
\tilde g_{\mu\nu} = \Omega^2 g_{\mu\nu}
}
where $\Omega$ is the conformal factor, and $g_{\mu\nu}$ here denotes the metric in the Jordan frame. 
Note that \Eref{conf_transf_gen} is equivalent to a transformation of units~\cite{ConfTransf}. 

In this section we review the field equations of $f(R)$ gravity in both the Jordan and Einstein frames, as well  
as those of the BD-equivalent form of $f(R)$ gravity\footnote{For various Hamiltonian formulations of $f(R)$ gravity see \cite{Deruelle2009}.}.

\subsection{Jordan Frame}

The action \eref{fofRaction} is called the Jordan frame action. 
An action is said to be in the Jordan frame, if the dynamical metric tensor in the action is the metric whose geodesics particles follow, i.e, the 
physical metric. The Jordan frame is the one in which the definition of the matter stress-energy tensor is
\labeq{1}{
T_{\mu\nu}^{(m)} = -\frac{2}{\sqrt{-g}}\frac{\delta S_m}{\delta g^{\mu\nu}},
}
where $\delta S_m/\delta g^{\mu\nu}$ is the functional derivative of $S_m$ with respect to $g^{\mu\nu}$.
For example, it is in this frame that the stress-energy tensor of a perfect fluid has the form
\labeq{2}{
T_{\mu\nu}^{(m)} = (\rho+ P)u_{\mu}u_{\nu} + P g_{\mu\nu},
}
where $\rho$ is the total energy density of the fluid, $P$ the fluid pressure, and $u^\mu$ the fluid four velocity.

Varying the action \eref{fofRaction} with respect to the metric
 yields the $f(R)$ field equations \cite{f_of_R_review,Sotiriou} in 
the Jordan frame
\labeq{fofREOM}{
\Sigma_{\mu\nu}=8\pi  T_{\mu\nu}^{(m)},
}
where $T_{\mu\nu}^{(m)}$ is the matter stress-energy tensor and
\labeq{Sigma}{
\Sigma_{\mu\nu}=FR_{\mu\nu}-\half f g_{\mu\nu}-\nabla_\mu\nabla_\nu F+g_{\mu\nu}\Box F,
}
and where $F = f,_R$. Note that for brevity we have dropped the argument of both $f(R)$ and $F(R)$. 
Clearly GR is recovered for $f = R - 2\Lambda$, in which case Equations \eref{fofREOM} and \eref{Sigma} yield
\labeq{3}{
\Sigma_{\mu\nu} = G_{\mu\nu}+\Lambda g_{\mu\nu} = 8\pi T_{\mu\nu}^{(m)},
}
where $G_{\mu\nu}$ is the Einstein tensor.

\Eref{fofREOM} is 4th-order due to the term $\nabla_\mu\nabla_\nu F$. However, if we take the trace of \Eref{fofREOM},
we obtain
\labeq{TrSigma}{
3\Box F + FR-2f = 8\pi T^{(m)},
}
where $T^{(m)} = g^{\mu\nu}T_{\mu\nu}^{(m)}$ and 
\labeq{4}{
\Box F = \frac{1}{\sqrt{- g}}\partial_\mu(\sqrt{- g} g^{\mu\nu}\partial_\nu F).
}
\Eref{TrSigma} can be used to promote $F(R)$ into an effective dynamical scalar
degree of freedom (often referred to as ``scalaron''), thus recasting the theory in 2nd-order form. 

\Eref{fofREOM} can also be written in the following form \cite{Starobinsky2}
\labeq{fofREOM2}{
G_{\mu\nu}= 8\pi (T_{\mu\nu}^{(m)}+T_{\mu\nu}^{(f)}),
}
where $T_{\mu\nu}^{(f)}$ can be thought of as a ``dark energy'' stress-energy tensor, given by
\beqar
8\pi T_{\mu\nu}^{(f)}= & &\ \half g_{\mu\nu}(f-R)+\nabla_\mu\nabla_\nu F \nonumber \\ 
                       & & -g_{\mu\nu}\Box F+ (1-F)R_{\mu\nu},
\eeqar

This form of the field equations of the theory is  interesting because the Bianchi identities
 $\nabla^\mu G_{\mu\nu}=0$ together with $\nabla^\mu T_{\mu\nu}^{(m)} = 0$, imply that 
\labeq{5}{
\nabla^\mu T_{\mu\nu}^{(f)} = 0,
}
i.e., the dark energy tensor $T_{\mu\nu}^{(f)} $ is conserved. 

\subsection{Einstein frame}

To obtain the Einstein frame action of $f(R)$ gravity, i.e., an action linear in a Ricci scalar $\tilde R$ associated with 
a metric $\tilde g_{\mu\nu}$, all we have to do is perform a conformal transformation on the metric 
\labeq{6}{
\tilde g_{\mu\nu} \equiv F g_{\mu\nu},
}
i.e., the conformal factor $\Omega$ in \Eref{conf_transf_gen} is $\Omega^2=F$.
For the transformation to be physical $F$ must satisfy $F>0$. Note that this condition is in accord with 
the first condition for cosmological viability of $f(R)$ gravity listed in \Sref{Introduction}.

If we introduce a new field $\phi$ such that
\labeq{7}{
\phi \equiv \sqrt{\frac{3}{16\pi}}\ln F,
}
then the $f(R)$ Jordan action transforms to \cite{f_of_R_review}
\labeq{fofREinstein}{
  S_E = \frac{1}{16\pi G}\int d^4x\sqrt{-\tilde g}\tilde R+S_\phi+S_m(F^{-1}(\phi)\tilde g_{\mu\nu},\psi_m),
}
where 
\labeq{8}{
S_\phi =\int d^4x\sqrt{-\tilde g}\big[-\half \tilde g^{\mu\nu}\partial_\mu\phi\partial_\nu\phi-V(\phi)\big]
}
is the scalar field term in the action, and where the scalar field potential is defined as
\labeq{9}{
V(\phi) = \frac{FR-f}{16\pi F^2}.
}

The dynamical metric tensor in the Einstein frame is not the physical ($g_{\mu\nu}$) but the conformal one ($\tilde g_{\mu\nu}$). However, 
the matter still follows the geodesics of the physical (Jordan) metric. 
Variation of the matter action with respect to $\tilde g_{\mu\nu}$ yields
\labeq{10}{
\tilde T_{\mu\nu}^{(m)} = -\frac{2}{\sqrt{-\tilde g}}\frac{\delta S_m}{\delta \tilde g^{\mu\nu}} = \frac{1}{F}T_{\mu\nu}^{(m)},
}
which is no longer independent of the scalar field $\phi$.

Variation of the action \eref{fofREinstein} with respect to $\phi$ yields the scalar field equation
\labeq{phi_einstein}{
 \tilde \Box \phi - V,_\phi - \sqrt{\frac{4\pi}{3}}\tilde T^{(m)} = 0,
}
where $\tilde T^{(m)} = \tilde g^{\mu\nu}\tilde T_{\mu\nu}^{(m)}$ and
\labeq{11}{
\tilde \Box \phi = \frac{1}{\sqrt{-\tilde g}}\partial_\mu(\sqrt{-\tilde g}\tilde g^{\mu\nu}\partial_\nu\phi).
}
\Eref{phi_einstein} implies that the scalar field is directly coupled to matter. 

Finally, variation of the action  \eref{fofREinstein}  with respect to $\tilde g^{\mu\nu}$ yields
\labeq{fofREinsteinEOM}{
\tilde G_{\mu\nu} = 8\pi (\tilde T_{\mu\nu}^{(m)}+\tilde T_{\mu\nu}^{(\phi)}),
}
where the scalar field stress-energy tensor is
\labeq{12}{
\tilde T_{\mu\nu}^{(\phi)}
                     =  \partial_\mu\phi\partial_\nu\phi-\tilde g_{\mu\nu}\bigg[\half\tilde g^{\alpha\beta}\ptl_\alpha\phi\ptl_\beta\phi+V(\phi)\bigg].
}

Note that in the Einstein frame $\tilde\nabla^\mu \tilde T_{\mu\nu}^{(m)} \neq 0$; instead we have
\labeq{13}{
  \tilde\nabla^\mu \tilde G_{\mu\nu} = \tilde\nabla^\mu (\tilde T_{\mu\nu}^{(m)}+\tilde T_{\mu\nu}^{(\phi)})= 0,
}
where $\tilde\nabla_\mu$ is the covariant derivative associated with $\tilde g_{\mu\nu}$. It can also be shown that \cite{f_of_R_review}
\labeq{14}{
\tilde\nabla^\mu \tilde T_{\mu\nu}^{(m)}= -\frac{1}{\sqrt{6}}\tilde T \tilde\nabla_\nu \phi, \qquad 
\tilde\nabla^\mu \tilde T_{\mu\nu}^{(\phi)}= \frac{1}{\sqrt{6}}\tilde T \tilde\nabla_\nu \phi.
}

\subsection{Equivalence with Brans-Dicke gravity}

Another way to cast $f(R)$ gravity into second-order form is to express the theory as a BD theory.
To show that $f(R)$ gravity is equivalent to BD gravity with a potential, the following action 
was considered in \cite{Chiba}:
\labeq{BDequiv}{
S =\frac{1}{16\pi } \int \sqrt{-g}d^4 x\big[f(\chi)+f,_\chi(\chi)(R-\chi)\big]+S_m.
}

Varying the action with respect to $\chi$ yields
\labeq{15}{
f,_{\chi\chi}(\chi)(R-\chi) = 0.
} 
Thus, if $f,_{\chi\chi}(\chi)\neq 0$ (in agreement with condition 3 in \Sref{Introduction}), then 
\labeq{BD1}{
\chi = R.
}
Hence, \Eref{BDequiv}
recovers the Jordan frame $f(R)$ action \eref{fofRaction}. If we now let $\phi = f,_{\chi}(\chi)$, \Eref{BDequiv} can be written as follows
\labeq{BDequiv2}{
  S =\frac{1}{16\pi} \int \sqrt{-g}d^4 x\bigg[\phi R - V(\phi)\bigg]+S_m,
}
where the potential is given by
\labeq{16}{
V(\phi) = \chi(\phi)\phi-f(\chi(\phi)).
}

Action \eref{BDequiv2} is the same as the original BD action with a potential and
 without the kinetic term  $(\omega/2) g^{\mu\nu}\ptl_\mu\phi\ptl_\nu\phi$, 
i.e., the BD parameter is $\omega = 0$. Varying the action \eref{BDequiv2} with respect to
 the metric yields the BD-equivalent $f(R)$ field equations \cite{Sotiriou}
\labeq{BD2}{
G_{\mu\nu} = \frac{8\pi }{\phi} (T_{\mu\nu}^{(m)}+T_{\mu\nu}^{(\phi)}),
}
where 
\labeq{TmnBDphi}{
8\pi  T_{\mu\nu}^{(\phi)} = \nabla_\mu\nabla_\nu\phi-g_{\mu\nu}\big(\Box\phi+\half V(\phi)\big)
}

 Taking the trace of \Eref{BD2} we can replace $R$ in \Eref{BD1} to obtain
\labeq{BD3}{
3\Box \phi + 2V(\phi)-\phi\frac{dV}{d\phi}=8\pi T.
}

Equations \eref{BD2} and \eref{BD3} are the BD-equivalent $f(R)$ field equations.

\section{Constraint Preservation in a spacetime context}
\label{con_pres}

In this section we review the concept
of constraint preservation using the standard Einstein equations in 4D covariant form, i.e., we do not invoke machinery 
of the 3+1 decomposition of spacetime. The reason for doing so is that so far we have written the most popular
 representations of metric $f(R)$ field equations in a GR-like form 
\labeq{GR}{
G_{\mu\nu} = 8\pi \bar T_{\mu\nu},
}
where $\bar T_{\mu\nu}$ is an ``effective'' stress-energy tensor that is conserved, i.e., $\nabla_\mu \bar T^{\mu\nu} =0$. 
Thus, it is instructive to first consider the Einstein equations in their familiar 4D covariant form.

For the Einstein equations with a cosmological constant we have $\bar T_{\mu\nu} = T_{\mu\nu}^{(m)}-\Lambda g_{\mu\nu}/8\pi$.
Since the Einstein equations are second-order partial differential equations, the evolution of the 4-metric $g_{\alpha\beta}$ in time can be determined
 by specifying $g_{\alpha\beta}$ and $\partial_t g_{\alpha\beta}$, everywhere on a three-dimensional spacelike hypersurface that
 corresponds to a given initial time $t$. \Eref{GR} can provide us with expressions for $\partial_t^2 g_{\alpha\beta}$, which we can use to 
 evolve the metric in time. There are 10 metric components and there are 10 field equations in~\eref{GR}. Hence, it appears that we have 
the exact number of equations for the 10 degrees of freedom of the metric.
 However, the Bianchi identities $\nabla_\beta G^{\alpha\beta} = 0$, give
\labeq{Bianchi_primit}{
\partial_t G^{\alpha 0} = -\partial_i G^{\alpha i} - G^{\beta\mu}{}^{(4)}\Gamma^{\alpha}{}_{\beta\mu}- G^{\alpha\beta}{}^{(4)}\Gamma^{\mu}{}_{\beta\mu},
}
where we set $\partial_t\equiv\partial_0$ and where ${}^{(4)}\Gamma^{\alpha}{}_{\beta\mu}$ are the Christoffel symbols associated with $g_{\alpha\beta}$.
Since no term on the right-hand-side of \Eref{Bianchi_primit} contains third time derivatives or higher, the four quantities $G^{\alpha 0}$ cannot
contain second time derivatives. Thus, the four equations 
\labeq{GR2}{
G_{\mu0} = 8\pi\bar T_{\mu0}
}
do not provide any information on the dynamical evolution of the metric. They are instead a set of constraints 
that $g_{\alpha\beta}$ and $\partial_t g_{\alpha\beta}$
have to satisfy. The only truly dynamical equations are the six remaining equations
\labeq{GR3}{
G_{ij} = 8\pi\bar T_{ij}.
}
The apparent mismatch between the number of metric components and the number of evolution equations is immediately resolved once we invoke the coordinate
freedom of GR. The theory is four-dimensional, and hence we can always choose four conditions to specify a coordinate system. For example, we can choose
the four $g_{0\beta}$ components and assign them certain values, or demand that they satisfy a given set of four partial differential equations. This way we are left
with six independent metric components, for which we have the exact number of evolution equations \eref{GR3}.

However, solving \Eref{GR3} does not guarantee that the full set of the Einstein equations \eref{GR} will be satisfied. For that to be true, 
\Eref{GR2} has to be satisfied for all times. In other words, if one solves \Eref{GR3} starting with initial data that satisfy \Eref{GR2}, 
one has to prove that the constraints are preserved. 

To demonstrate that this is indeed the case we make use of the Bianchi identities in the following form:
\labeq{Bianchi_primit2}{
\nabla_\beta {\mathcal E}^{\alpha\beta} = 0, 
}
or
\labeq{Bianchi_primit3}{
\partial_t \mathcal{E}^{\alpha 0} = -\partial_i \mathcal{E}^{\alpha i} - \mathcal{E}^{\beta\mu}{}^{(4)}\Gamma^{\alpha}{}_{\beta\mu}- \mathcal{E}^{\alpha\beta}{}^{(4)}\Gamma^{\mu}{}_{\beta\mu},
}
where
\labeq{17}{
{\mathcal E}^{\alpha\beta} \equiv G^{\alpha\beta} - 8\pi \bar T^{\alpha\beta}.
}
If we let $C = {\mathcal E}^{00}$ and $C^i={\mathcal E}^{i0}$, \Eref{Bianchi_primit3} can be rewritten as
\beqar \label{spacetime_constraint_prop}
\partial_t C =& &  -\partial_i C^i - C \big(2{}^{(4)}\Gamma^{0}{}_{00}+{}^{(4)}\Gamma^{i}{}_{0i}\big)\nonumber \\ 
               &  & -C^i\big(3{}^{(4)}\Gamma^{0}{}_{i0}+{}^{(4)}\Gamma^{j}{}_{ij}\big) \\
\partial_t C^j  = & &  -C{}^{(4)}\Gamma^{j}{}_{00}-2C^i{}^{(4)}\Gamma^{j}{}_{i0}-C^j{}^{(4)}\Gamma^{\beta}{}_{0\beta},
\eeqar
where we have used \Eref{GR3}, ${\mathcal E}^{ij}=0$, to obtain the result. Thus, if the constraints are initially satisfied, then $C=C^i=0$ initially 
and from \Eref{spacetime_constraint_prop} the time derivative of the constraints will be zero and hence the constraints 
will remain zero for all times. Since this conclusion resulted from setting ${\mathcal E}^{ij}=0$, 
the previous statement is equivalent to saying that the evolution equations preserve the constraints, 
a result that is well-known.

\section{3+1 Decomposition of $f(R)$ gravity}
\label{3p1_intro}

Well-posedness of the Cauchy problem in metric $f(R)$ gravity has been demonstrated in \cite{fofR_well_posed} using the BD-equivalent $f(R)$ formulation.
In this section we focus on the BD version of $f(R)$ gravity and review the salient features of its 3+1 decomposition that will be useful
in our proof of constraint preservation.

The form of the field equations of the theory is that of \Eref{GR}, where 
\labeq{18}{
\bar T_{\mu\nu} = \frac{1}{\phi} (T_{\mu\nu}^{(m)}+T_{\mu\nu}^{(\phi)}).
}

The 3+1 decomposition of spacetime is a decomposition of spacetime into space and time. To do this, one assumes that the 
four-dimensional spacetime manifold can be foliated by a one-parameter family of nonintersecting 
spacelike hypersurfaces. The parameter of this family of hypersurfaces is taken to be the coordinate time.
The spacetime metric is then rewritten as \cite{ADM3p1}
\begin{equation}
ds^2 = -\alpha^2 dt^2 + \gamma_{ij} (dx^i + \beta^i dt) (dx^j + \beta^j dt),
\end {equation}
where $\alpha$ is the lapse function, $\beta^i$ is the shift vector, and $\gamma_{ij}$ is the 3-metric on the spacelike 
hypersurfaces, induced by $g_{\alpha\beta}$. The lapse function and the shift vector are gauge quantities; they
dictate how to build the coordinate system and can be freely specified.
The relation between $\gamma_{ij}$ and $g_{\alpha\beta}$ is
\labeq{projection_op}{
\gamma^{\alpha}{}_{\beta} = \delta^{\alpha}{}_{\beta} + n^\alpha n_{\beta},
}
where $\gamma^{\alpha}{}_{\beta}= g^{\alpha\mu}\gamma_{\mu\beta}$, $\delta^{\alpha}{}_{\beta}$ is the Kronecker delta, and
$n^{\alpha}$ is the future directed timelike unit vector normal to the $t = \rm const.$ hypersurfaces.
The tensor $\gamma^{\alpha}{}_{\beta}$ is the operator that projects tensors onto spacelike hypersurfaces.

The field equations  can then be decomposed into a set of evolution equations and a set of constraint equations
by using $\gamma^{\alpha}{}_{\beta}$ and $n^{\alpha}$.

Projecting \Eref{GR} twice with the projection operator yields the evolution equations
\labeq{evolution_prim}{
E_{\mu\nu} \equiv (G_{\alpha\beta} - 8\pi \bar T_{\alpha\beta}) \gamma^\alpha{}_\mu \gamma^\beta{}_\nu = 
G_{\alpha\beta} \gamma^\alpha_\mu \gamma^\beta_\nu - 8\pi \bar S_{\mu\nu}=0,
}
where 
\labeq{19}{
\bar S_{\mu\nu} \equiv \bar T_{\alpha\beta}\gamma^\alpha{}_\mu \gamma^\beta{}_\nu = S_{\mu\nu} + S_{\mu\nu}^{(\phi)},
}
and where 
\labeq{Smn}{
S_{\mu\nu} \equiv T_{\alpha\beta}^{(m)}\gamma^\alpha{}_\mu \gamma^\beta{}_\nu, \qquad 
S_{\mu\nu}^{(\phi)} \equiv T_{\alpha\beta}^{(\phi)}\gamma^\alpha{}_\mu \gamma^\beta{}_\nu.
}
Using \Eref{TmnBDphi}, we can write $S_{\mu\nu}^{(\phi)}$ in \Eref{Smn} as follows
\labeq{20}{
S_{\mu\nu}^{(\phi)} = \frac{1}{8\pi}[D_\mu\nabla_\nu \phi - \gamma_{\mu\nu}(\Box \phi +\half V(\phi))], 
}
where $D_\mu$ is the covariant derivative associated with $\gamma_{\mu\nu}$.

Contracting \Eref{GR} twice with $n^{\alpha}$ yields the Hamiltonian constraint
\labeq{Hamiltonian_prim}{
H \equiv (G_{\alpha\beta} - 8\pi \bar T_{\alpha\beta}) n^\alpha n^\beta = 
G_{\alpha\beta}n^\alpha n^\beta - 8\pi \bar \rho = 0,
}
where 
\labeq{21}{
\bar \rho \equiv \bar T_{\alpha\beta}n^\alpha n^\beta = \rho + \rho^{(\phi)},
}
and where 
\labeq{22}{
\rho \equiv T_{\alpha\beta}^{(m)}n^\alpha n^\beta, \qquad \rho^{(\phi)} \equiv T_{\alpha\beta}^{(\phi)}n^\alpha n^\beta
}

Using \Eref{TmnBDphi} we also obtain
\labeq{23}{
\rho^{(\phi)} = \frac{1}{8\pi} [n^{\mu}n^{\nu}\nabla_\mu\nabla_\nu\phi + (\Box \phi +\half V(\phi))].
}

Contracting \Eref{GR} once with $n^{\alpha}$ and projecting once with $\gamma^\alpha{}_\beta$ yields the momentum constraints
\beqar
\label{Momentum_prim}
M_\mu & \equiv & -(G_{\alpha\beta} - 8\pi \bar T_{\alpha\beta}) n^\alpha \gamma^\beta{}_\mu \nonumber \\ 
        &  =   &  -G_{\alpha\beta}n^\alpha \gamma^\beta{}_\mu - 8\pi j_{\mu} - 8\pi j_{\mu}^{(\phi)}=0 ,
\eeqar
where 
\labeq{24}{
j_\mu \equiv - T_{\alpha\beta}^{(m)}n^\alpha \gamma^\beta{}_\mu, \qquad j_\mu^{(\phi)} \equiv - T_{\alpha\beta}^{(\phi)}n^\alpha \gamma^\beta{}_\mu,
}
and where from \Eref{TmnBDphi} we find
\labeq{25}{
j_\mu^{(\phi)} = \frac{1}{8\pi}n^\alpha\gamma^\beta{}_\mu \nabla_\alpha\nabla_\beta\phi.
}


We can now write \Eref{GR} as a linear combination of the evolution and the constraint equations.
\begin{eqnarray}
G^{\alpha\beta} - 8\pi \bar T^{\alpha\beta} &=& (G^{\mu\nu} - 8\pi \bar T^{\mu\nu}) \delta^\alpha{}_\mu \delta^\beta{}_\nu \nonumber  \\ 
                                            &=& (G^{\mu\nu} - 8\pi \bar T^{\mu\nu})( \gamma^\alpha{}_\mu - n^\alpha n_\mu)( \gamma^\beta{}_\nu - n^\beta n_\nu)
\label{Ein_decomp0}
\end{eqnarray}
where we used \Eref{projection_op} in the second line to replace the Kronecker deltas. By use of Equations \eref{evolution_prim}, \eref{Hamiltonian_prim} and \eref{Momentum_prim}, \Eref{Ein_decomp0} becomes
\labeq{Ein}{
 G^{\alpha\beta} - 8\pi \bar T^{\alpha\beta} = E^{\alpha\beta} + 2 n^{(\alpha} M^{\beta)} + H n^\alpha n^\beta.
}
This last equation has been derived by Frittelli, (see Eq. (9) in \cite{Frit97}) for GR. Here we have shown that this equation is valid 
in $f(R)$ gravity, too, provided that appropriate definitions of $H$ and $M^i$ are given.

As was shown in \cite{Frit97}, setting $E^{\alpha\beta} = 0$ yields the evolution equations of the original ADM formulation
\cite{ADM3p1}, whereas setting $E^{\alpha\beta} = \gamma^{\alpha\beta}H$ yields the evolution equations of the standard ADM formulation 
\cite{ADMbyYork,AlcuBook,BSBook}. Following the parametrization of \cite{Frit97} we set $E^{\alpha\beta} = \lambda\gamma^{\alpha\beta}H$, 
so that in the ADM language $\lambda=0$ corresponds to the
original ADM formulation, while $\lambda=1$ to the standard ADM formulation, except that here we deal with $f(R)$ 3+1 formulations.

It is now evident from \Eref{Ein} that if $M^\alpha = H = 0$ and $E^{\alpha\beta} = \lambda\gamma^{\alpha\beta}H$, 
then the $f(R)$ equations are satisfied.

By introducing the extrinsic curvature $K_{ij}$
\labeq{Kij}{
K_{ij} = -\frac{1}{2}\pounds_n \gamma_{ij}
} 
where $\pounds_n$ stands for the Lie derivative along the timelike unit vector $n^\alpha$,
using the Gauss, Godazzi and Ricci equations (see e.g. Equations (2.68), (2.73), (2.82) in \cite{BSBook}), and adopting the usual coordinate basis where
\labeq{nmu}{
n^\mu = (\alpha^{-1}, - \alpha^{-1} \beta^i) 
}
one can derive the evolution and constraint equations in 3+1 form, which (for $\lambda=0$) are presented in \cite{fofR_well_posed} and we do not 
repeat them here. 

A subtlety that must be addressed for our purpose and which is pointed out in \cite{Salgado06,fofR_well_posed}
is that to remove the time derivatives of the scalar field
$\phi$ from the sources $S_{\mu\nu}^{(\phi)},\rho^{(\phi)},j_{\mu}^{(\phi)}$ one introduces the gradients of $\phi$ as new dynamical variables
\labeq{Pi}{
\Pi \equiv \pounds_n\phi= n^\mu \nabla_\mu \phi,
}
\labeq{Qmu}{
Q_\mu \equiv D_\mu \phi.
}

The $\Box \phi$ operator in the sources $S_{\mu\nu}^{(\phi)},\rho^{(\phi)},j_{\mu}^{(\phi)}$ can be removed by use of \Eref{BD3}. 
Furthermore, \Eref{BD3} in combination with \Eref{Pi}, which can be written as
\labeq{Pievol}{
\alpha\Pi = \partial_t \phi - \beta^i Q_i,
}
can be used to derive the evolution equation for $\Pi$. Eventually, one finds \cite{Salgado06}
\labeq{26}{
\pounds_{n} \Pi = \Pi K + Q^iD_i(\ln\alpha)+D_i Q^i-\Box \phi.
}
where $K=\gamma^{ij}K_{ij}$.

To promote $Q_i$ to a dynamical variable we take a time derivative of $Q_i$ and using \Eref{Pievol} we obtain 
\labeq{Qievol}{
\partial_t Q_i = \pounds_\beta Q_i + D_i (\alpha\Pi),
}
where 
\labeq{27}{
\pounds_\beta Q_i = \beta^s\ptl_s Q_i + Q_s \ptl_i \beta^s.
}

The introduction of new variables $Q_i$, introduces an extra constraint, which the evolution equations have to satisfy
\labeq{Cmu}{
C_{i} \equiv Q_{i} - D_i \phi = 0.
}
 In addition to this, the ordering constraint 
\labeq{Cmunu}{
C_{ij} \equiv D_i Q_j - D_j Q_i = 0.
}
has to be satisfied, too.

Thus, constraint preservation means that the evolution equations must preserve all the constraints of the 3+1 decomposition, i.e., 
Equations \eref{Hamiltonian_prim},~\eref{Momentum_prim},~\eref{Cmu}, and~\eref{Cmunu}.\\


\section{Constraint Propagation Equations of 3+1 $f(R)$ gravity}
\label{General_Con_prop}

The backbone of the Frittelli approach is to express the field equations in the form of \Eref{Ein} and plug it in the Bianchi 
identities in order to derive the evolution equations for the constraints, assuming the evolution equations are satisfied $E^{\mu\nu}=\lambda \gamma^{\mu\nu}H$. 
So far we have extended the Frittelli approach
 to general metric $f(R)$ gravity. Since the form of Equations \eref{Ein} is the same as in \cite{Frit97},   
the derivation of the 3+1 BD-equivalent $f(R)$ constraint propagation equations
is precisely the same as that in \cite{Frit97}, which is valid for GR, and to which we refer the interested reader for more details. 
Here we only sketch the derivation and write the result.

The Bianchi identities are
\labeq{Bianchi}{
\nabla_\mu (G^{\mu\nu}-8\pi \bar T^{\mu\nu}) = 0.
}
or, equivalently, after substituting \Eref{Ein} in \Eref{Bianchi}
\labeq{Bianchi2}{
\nabla_\mu (E^{\mu\nu} + 2 n^{(\mu} M^{\nu)} + H n^\mu n^\nu) = 0.
}

To find the evolution of the Hamiltonian constraint we contract \Eref{Bianchi2} with $n^\alpha$ 
and after some algebra we find

\begin{eqnarray} \label{Ham_evol_Frit}
0 & = & -E^{\mu\nu} D_\nu n_\mu - 2 n^\nu M^\mu \nabla_\nu n_\mu - D_\nu M^\nu \nonumber \\ 
 & &  -\ n^\nu \nabla_\nu H - H D_\nu n^\nu.
\end{eqnarray}

To find the evolution of the Momentum constraint we project \Eref{Bianchi2} with $\gamma^\alpha{}_\beta$ 
and after some algebra we find \footnote{We note here that Eq. (11) in \cite{Frit97}, differs from our \Eref{Mom_evol_Frit} 
by a factor of 2 in the term $ H n^\mu \nabla_\mu n^\alpha$. 
We believe that this discrepancy is simply due to a typographical error.\label{ftnt1}}

\begin{eqnarray} 
 0  & = &  D_\mu E^{\mu \alpha} + n^\nu E^{\alpha\mu} \nabla_\nu n_\mu +  n^\mu \nabla_\mu M^\alpha - n^\alpha M^\nu n^\mu \nabla_\mu n_\nu \nonumber  \\
     & &  +\ M^\alpha D_\mu n^\mu + M^\mu D_\mu n^\alpha + H n^\mu \nabla_\mu n^\alpha. \label{Mom_evol_Frit}
\end{eqnarray}

Proof that our equations are correct will be provided below when we cast the constraint propagation equations in pure 3+1
language and compare our result with published results in the literature obtained via the 3+1 approach.

Using the following identities
\labeq{28}{\gamma^{\mu\nu} H D_\mu n_\nu = H D_\mu n^\mu,}
\labeq{29}{n^\mu \gamma^{\alpha\nu} H \nabla_\mu n_\nu =   H n^\mu \nabla_\mu n^\alpha,}
and substituting $E^{\mu\nu}=\lambda\gamma^{\mu\nu}H$ in Equations~\eref{Ham_evol_Frit} and~\eref{Mom_evol_Frit}
we  find that the evolution of the constraints is given by
\labeq{con_prop_ham}{
 n^\mu \nabla_\mu H  = - 2 n^\mu M^\nu \nabla_\mu n_\nu - D_\mu M^\mu - (1+\lambda) H D_\mu n^\mu,
}  
\beqar\label{con_prop_mom}
n^\mu \nabla_\mu M^\nu = & & -\lambda \gamma^{\mu\nu} D_\mu H + n^\nu M^\alpha n^\beta \nabla_\beta n_\alpha - M^\nu D_\mu n^\mu \nonumber \\
                  & & - M^\mu D_\mu n^\nu -  (1+\lambda) H n^\mu \nabla_\mu n^\nu.
\eeqar

These last two equations have the same mathematical form (except for a factor of 2; see footnote in page \pageref{ftnt1}) 
 as those derived in \cite{Frit97} that applied to the case of GR, i.e. $f(R)=R$.
Here we have proven that the form of the hamiltonian and momentum constraint
 propagation equations is the same for both vacuum ($T_{\mu\nu}^{(m)} = 0$) and non-vacuum spacetimes ($T_{\mu\nu}^{(m)} \neq 0$),
and that it is independent of the $f(R)$ function, because we have absorbed all terms that depend on these quantities
in the definition of the hamiltonian and momentum constraints  
(see Equations~\eref{Hamiltonian_prim},~\eref{Momentum_prim}). 

We deal with the evolution of constraints \eref{Cmu} and \eref{Cmunu}, in the following section.

\section{$f(R)$ Constraint propagation equations in pure 3+1 language}
\label{3p1_lang_con_prog}

Note that Equations~\eref{con_prop_ham} and~\eref{con_prop_mom} involve both spacetime and purely spatial 
objects. This is not a form that easily yields a comparison between the constraint propagation equations obtained
in the Frittelli approach with those obtained in the 3+1 approach. Nor is it convenient for 
integration in a 3+1 numerical implementation that could serve as a check of the numerical integration of the evolution equations of the 
dynamical variables. 
For this reason, we now cast these equations
 in pure 3+1 language. To our knowledge such a calculation has never been published before, hence it is instructive to include it 
here.


Alternative expressions for the extrinsic curvature are (see e.g. Equations~(2.49), (2.52) in  \cite{BSBook})
\labeq{ext_curv_2}{D_{(\alpha} n_{\beta)} = - K_{\alpha\beta} \mbox{\ \ and\ \ } K_{\alpha\beta} = -\nabla_\alpha n_\beta  - n_\alpha a_\beta, }
where $a_\alpha = n^\beta\nabla_\beta n_\alpha$ is the acceleration of normal observers, also equal to (see Eq.~(2.22) in \cite{BSReview})
\labeq{accel}{
a_\beta = D_\beta \ln\alpha.
}

From \Eref{ext_curv_2} it can be shown that 
\labeq{ext_curv_sca}{D_\beta n^\beta = -K \mbox{\ \ and\ \ } D_\beta n^\alpha = - K_\beta{}^\alpha.} 
By use of Equations~\eref{ext_curv_2},~\eref{accel} and~\eref{ext_curv_sca}, Equations~\eref{con_prop_ham} and~\eref{con_prop_mom}
can be written as 
\begin {eqnarray}
 n^\mu \nabla_\mu H  &=&    - D_\mu M^\mu + (1+ \lambda) H K  \nonumber  \\
                         & &  - 2 M^\nu D_\nu \ln \alpha  \label{Ham_con_new}, \\
 n^\mu \nabla_\mu M^\nu &=&  -\mu \gamma^{\mu\nu} D_\mu H +  n^\nu M^\mu D_\mu \ln \alpha + M^\nu K \nonumber \\
                                   & &  + M^\mu K_\mu{}^\nu - (1+ \lambda) H D^\nu \ln \alpha. \label{Mom_con_new} 
\end{eqnarray}
The identities, $\nabla_\alpha H =  \partial_\alpha H$, $\gamma^{\mu\nu}D_\mu H=\gamma^{\mu\nu}\partial_\mu H$, 
$M^{\alpha}D_\alpha \ln \alpha = M^{\alpha}\partial_\alpha \ln \alpha$,
 $\nabla_\mu M^\nu = \partial_\mu M^\nu + {}^{(4)}\Gamma^\nu_{\mu\beta} M^\beta$, can be
 used to replace the covariant derivatives that occur above. 
Furthermore,  the timelike unit vector ($n^\mu$) can be replaced by \Eref{nmu}. 
Equations~\eref{Ham_con_new}
and~\eref{Mom_con_new} can then be written as
\labeq{Ham_con_new2}{
\partial_t H  =  \beta^i \partial_i H - 2 M^i \partial_i \alpha - \alpha D_i M^i +(1+ \mu) \alpha H K,
}
%
\beqar\label{Mom_con_new2}
\partial_t M^j = & & -\mu \gamma^{ij} \partial_i H + \beta^i \partial_i M^j - {}^{(4)}\Gamma^j_{i0} M^i \nonumber \\ 
                 & & + {}^{(4)}\Gamma^j_{ik} M^i \beta^k  + n^j M^i \partial_i \alpha + \alpha M^j K \nonumber \\ 
                 & & + \alpha M^i K_i^j - (1+ \mu) \gamma^{ij} H \partial_i \alpha,
\eeqar
where we have focused on the spatial indices of $M^\mu$, since $M^\mu$ is purely spatial.

Using $D_i M^i=\gamma^{ij}\partial_i M_j+\gamma^{ij}\Gamma^{k}{}_{ij}M_k$, 
 and the expressions for the Lie derivatives of the constraints along $\alpha n^\mu$
\begin{eqnarray}
\pounds_{\alpha n} H &=& \partial_t H - \beta^i \partial_i H \label{eq:old42}, \\
\pounds_{\alpha n} M^j &=& \partial_t M^j - \beta^i \partial_i M^j + M^i \partial_i \beta^j \label{eq:old43}
\end{eqnarray}
we write Equations~\eref{Ham_con_new2} and ~\eref{Mom_con_new2} as

\beqar\label{H_evol3}
\pounds_{\alpha n} H = & &  - 2 M^i \partial_i \alpha - \alpha \gamma^{ij}\partial_i M_j \nonumber \\ 
                       & & + \alpha\gamma^{ij}\Gamma^{k}{}_{ij}M_k +(1+ \mu) \alpha H K, 
\eeqar
\begin{eqnarray}
\pounds_{\alpha n} M^j = & & -\lambda \gamma^{ij} \partial_i H + A^j{}_i M^i + \alpha M^j K \nonumber\\ 
                  & & - (1+ \lambda) \gamma^{ij} H \partial_i \alpha+ \gamma^{jk} M^i \partial_i \beta_k \nonumber \\
                 & & - M^\ell\beta^s\gamma^{jm}\partial_\ell \gamma_{sm}, \label{M_evolb}
\end{eqnarray}
where 
\labeq{Aji_mom}{
A^j{}_i\equiv - {}^{(4)}\Gamma^j_{i0} + {}^{(4)}\Gamma^j_{ik} \beta^k -\alpha^{-1}\beta ^j\partial_i\alpha +\alpha K_i^j.
}

We now need to express the Christoffel symbols associated with the spacetime metric $g_{\mu\nu}$, that appear in \Eref{Aji_mom}, in terms
of the 3-metric $\gamma_{ij}$ and the gauge variables. We do this as follows

\beqar\label{Gamma0}
{}^{(4)}\Gamma^{j}{}_{i0} & = & \ \half g^{j\rho}(\partial_i g_{0\rho}+\partial_0 g_{i\rho}-\partial_\rho g_{i0}) \nonumber \\
                          & = & \ \half g^{j0}\partial_i g_{00} + \half g^{j\ell}(\partial_i g_{0\ell}+\partial_0 g_{i\ell}-\partial_\ell g_{i0}). 
\eeqar
Using the relations between $g_{\mu\nu}$ and $\alpha, \beta^i$, $\gamma_{ij}$ \cite{BSReview}
\labeq{ident}{
g_{00} = -\alpha^2 + \beta_\ell \beta^\ell, \quad g_{0i} = \beta_i, \quad g^{j\ell} = \gamma^{j\ell} - \alpha^{-2}\beta^j\beta^\ell,
}
\Eref{Gamma0} eventually becomes
\beqar\label{Gamma0_2}
{}^{(4)}\Gamma^{j}{}_{i0} =& & -\alpha^{-1}\beta^j\partial_i \alpha + \half\alpha^{-2}\beta^j\beta^\ell\partial_i\beta_\ell \nonumber \\
                           &  & -\half\big(\alpha^{-2}\beta^j\beta^\ell\beta^s\partial_i\gamma_{\ell s} - \gamma^{j\ell}\partial_i\beta_\ell
                              -\gamma^{j\ell}\partial_0\gamma_{i\ell}\big) \nonumber \\
			   &  & -\half\big(\gamma^{j\ell}\partial_\ell \beta_i +\alpha^{-2}\beta^j\beta^\ell\partial_0 \gamma_{i\ell}-
                              \alpha^{-2}\beta^j\beta^\ell\partial_\ell \beta_i\big), 
\eeqar
where we have also used the following identities
\labeq{ident2}{
\beta^i = \gamma^{ij}\beta_j, \quad \partial_k\gamma^{ij} = -\gamma^{is}\gamma^{jm}\partial_k \gamma_{sm}.
}

The next object that appears in \Eref{Aji_mom}, and which we cast in 3+1 language is $\beta^k {}^{(4)}\Gamma^{j}{}_{ik}$. We can write this as
\beqar\label{betaGamma}
\beta^k {}^{(4)}\Gamma^{j}{}_{ik} & = &\ \half \beta^k g^{j\rho}(\partial_i g_{k\rho}+\partial_k g_{i\rho} - \partial_\rho g_{ik})  \nonumber \\
                                  & = &\ \half \beta^k g^{j0}(\partial_i g_{k0}+\partial_k g_{i0} - \partial_0 g_{ik})  \nonumber \\
                            & & +  \half \beta^k g^{j\ell}(\partial_i g_{k\ell}+\partial_k g_{i\ell} - \partial_\ell g_{ik}).  
\eeqar

By virtue of Equations~\eref{ident} and \eref{ident2}, \Eref{betaGamma} finally becomes
\beqar\label{betaGamma2}
\beta^k {}^{(4)}\Gamma^{j}{}_{ik} = & &\ \half\alpha^{-2} \beta^k\beta^j\partial_i\beta_k + \half\alpha^{-2} \beta^k\beta^j\partial_k\beta_i \nonumber \\ 
                                  & & -\half\alpha^{-2} \beta^k\beta^j\partial_0\gamma_{ik} +\beta^{k}\Gamma^{j}{}_{ik} \nonumber \\ 
                                  & & -\alpha^{-2}\beta^j\beta^k\beta_\ell\Gamma^{\ell}{}_{ik}, 
\eeqar
or equivalently
\beqar\label{betaGamma3}
\beta^k {}^{(4)}\Gamma^{j}{}_{ik} = & &\ \half\alpha^{-2} \beta^k\beta^j\partial_i\beta_k + \half\alpha^{-2} \beta^k\beta^j\partial_k\beta_i \nonumber \\ 
                                  & & - \half\alpha^{-2} \beta^k\beta^j\partial_0\gamma_{ik} +\beta^{k}\Gamma^{j}{}_{ik} \nonumber \\ 
                                  & &  -\half\alpha^{-2}\beta^j\beta^k\beta^\ell\partial_i\gamma_{k\ell}, 
\eeqar
where $\Gamma^{j}{}_{ik}$ stand for the Christoffel symbols associated with the 3-metric. 

By use of Equations \eref{Gamma0_2} and \eref{betaGamma3}, \Eref{Aji_mom} becomes
%
\beqar\label{Aij2}
A^{j}{}_{i} = & & -\half \gamma^{j\ell}\partial_i\beta_\ell-\half\gamma^{j\ell}\partial_0\gamma_{i\ell}+
               \half\gamma^{j\ell}\partial_\ell\beta_i \nonumber \\
             & & +\beta^{k}\Gamma^{j}{}_{ik} + \alpha K^{j}{}_{i}.
\eeqar

From the evolution equation of the 3-metric, \Eref{Kij}, we have

\labeq{gam_dt_gam}{
\half \gamma^{j\ell}\partial_0\gamma_{i\ell} = -\alpha K^j{}_{i}+\half \gamma^{j\ell}\partial_i\beta_\ell+\half \gamma^{j\ell}\partial_\ell\beta_i
                                                -\gamma^{jl}\Gamma^s{}_{i\ell}\beta_s.
}

Substitution of \Eref{gam_dt_gam} into \Eref{Aij2} yields
\labeq{Aji3}{
A^j{}_i = -\gamma^{j\ell}\partial_i\beta_\ell + \gamma^{j\ell}\beta^k\partial_i\gamma_{\ell k}+2\alpha K^j{}_i.
}

Finally, substituting \Eref{Aji3} into \Eref{M_evolb} yields the desired result,
\beqar\label{Mupj_evol}
\pounds_{\alpha n} M^j = & &-\lambda \gamma^{ij} \partial_i H + 2\alpha K^j_i M^i + \alpha M^j K \nonumber \\ 
                        & & - (1+ \lambda) \gamma^{ij} H \partial_i \alpha. 
\eeqar

Equations \eref{H_evol3} and \eref{Mupj_evol} are the hamiltonian and momentum constraint propagation equations in pure 3+1 language, where the Lie
derivatives are given in Equations~\eref{eq:old42} and~\eref{eq:old43}.

We have already established that the form of the constraint propagation equations is the same for both vacuum and non-vacuum spacetimes,
and is independent of the form of the function $f(R)$. 
Thus, to validate our equations we can use known results that apply to the case of GR, and have been derived
 by using the ``brute force'' method.

For this reason we now compare our results with results published in \cite{VasPas_formulations,YonShin2} that apply for $f(R)=R$, i.e., for the Einstein equations. 
 In these two papers the evolution equations of the constraints were presented 
assuming $T_{\mu\nu}=0$. In \cite{VasPas_formulations} the 3+1 approach was employed to derive
the constraint propagation equations. For direct comparison with these published results we also derive the evolution
 equations for $\mathcal{H} = 2  H$ and the evolution for $M_i$ which were used in \cite{VasPas_formulations,YonShin2}
 instead. 

Using
\beqar\label{Mlowj_evol}
\pounds_{\alpha n} M_i & = & M^j\pounds_{\alpha n}\gamma_{ij} + \gamma_{ij} \pounds_{\alpha n} M^j \nonumber \\
                       & = & - 2\alpha K_{ij} M^j + \gamma_{ij} \pounds_{\alpha n} M^j
\eeqar
and replacing $H = \mathcal{H}/2$ in  \eref{H_evol3} and \eref{Mupj_evol} we obtain the following alternative form for the constraint propagation equations 
%
\beqar\label{H_evol_final}
\pounds_{\alpha n} \mathcal{H} = & & - 4 M^i \partial_i \alpha - 2\alpha \gamma^{ij}\partial_i M_j \nonumber \\ 
                       & & + 2\alpha\gamma^{ij}\Gamma^{k}{}_{ij}M_k + (1+ \lambda) \alpha \mathcal{H} K ,
\eeqar
\labeq{Mlowj_evol_final}{
\pounds_{\alpha n} M_i = -\half\lambda \partial_i \mathcal{H}  + \alpha M_i K  - (1+ \lambda)\half \mathcal{H} \partial_i \alpha. 
}
For $\lambda=1$ Equations ~\eref{H_evol_final} and~\eref{Mlowj_evol_final} become precisely the same as the expressions
 in \cite{VasPas_formulations}, when the quantities, $C_{kij}=\partial_k\gamma_{ij}-D_{kij}$, defined in \cite{VasPas_formulations}, satisfy
$C_{kij}=0$. In that work $C_{kij}$ are constraints that arise from the introduction of the auxiliary variables
$D_{kij}\equiv\partial_k\gamma_{ij}$, which were used to reduce the ADM formulation to 1st-order. 
Also, a straightforward calculation shows that the expressions above 
are equivalent to the corresponding expressions in \cite{YonShin2}. From Equations~\eref{H_evol_final} and~\eref{Mlowj_evol_final} it is again 
evident that the constraints remain satisfied ($\mathcal{H}=M_i=0$), if they are initially satisfied.

We now turn our attention to the evolution equations of $C_{i}$ and $C_{ij}$. To derive the evolution of $C_{i}$ and $C_{ij}$ we simply take a 
time derivative of $C_{i}$ and $C_{ij}$, use the commutation relation $\partial_i \partial_t = \partial_t\partial_i$, and 
replace the time derivative of variables via Equations \eref{Pievol} and \eref{Qievol} to find that
\labeq{Cievol}{
\partial_t C_i = \beta^sC_{si}
}
and
\labeq{Cijevol}{
\partial_t C_{ij} = \pounds_\beta C_{ij}, 
}
where 
\labeq{30}{
\pounds_\beta C_{ij} = \beta^s\partial_s C_{ij} + C_{sj}\ptl_i\beta^s + C_{is}\ptl_j\beta^s.
}

Equations \eref{Cievol} and \eref{Cijevol} imply that if the constraints $C_i,C_{ij}$ are initially satisfied, 
then the evolution equations will preserve the constraints. 

We stress again that Equations~\eref{H_evol_final},~\eref{Mlowj_evol_final}, \eref{Cievol} and \eref{Cijevol} are valid not only for vacuum, 
but also for non-vacuum spacetimes, as well as for any viable $f(R)$ model. This is a new result that to our knowledge has not
 been pointed out previously and is not trivial to prove, if one
 employs the 3+1 or ``brute force'' method to derive the constraint propagation equations. Here we proved this without prior 
knowledge of the evolution equations of the dynamical variables $K_{ij}, \gamma_{ij}$, on the basis of the Frittelli approach. 

Finally, we note that  
the agreement between our expressions and results obtained by the ``brute force'' approach confirms that 
\Eref{Mom_evol_Frit} is correct (see footnote in page \pageref{ftnt1}).

\section{Constraint Propagation in the Jordan and Einstein frames}
\label{jordan_Einstein}

The form of the $f(R)$ field equations both in the Jordan frame (see Equations \eref{TrSigma} and \eref{fofREOM2}) and in the Einstein frame
(see Equations \eref{phi_einstein} and \eref{fofREinsteinEOM}) is the same as the BD formulation of $f(R)$ gravity. For this reason, it is 
evident from our discussion in \Sref{General_Con_prop} that the form of the constraint propagation equations in these two
frames must be the same as those in the BD formulation, provided that we define the hamiltonian and momentum constraints
analogously to Equations \eref{Hamiltonian_prim} and \eref{Momentum_prim} with one important caveat; 
The foliation in the Einstein frame must be based on the conformal (Einstein) metric and not the 
physical (Jordan), i.e., the induced 3-metric on spacelike hypersurfaces must be $\tilde \gamma_{ab} = \tilde g_{ab} + \tilde n_a \tilde n_b$, where
the normal timelike vector now satisfies $\tilde g_{ab}\tilde n^a \tilde n^b = -1$ and not $g_{ab}\tilde n^a \tilde n^b = -1$. Note that this last condition
is only a mathematical requirement for the 3+1 machinery to remain the same. Physical conclusions must still be drawn based on the Jordan metric.

Finally, we note that if one applies the general recipe for a 3+1 decomposition (see \Sref{3p1_intro}) to 
more general scalar-tensor theories of gravity considered in \cite{fofR_well_posed}, then 
the constraint propagation equations will be the same as our Equations~\eref{H_evol_final},~\eref{Mlowj_evol_final}, 
\eref{Cievol} and \eref{Cijevol}. This is because the covariant (Jordan frame) formulation of these theories obtains 
the same form as Equations~\eref{BD2} and~\eref{BD3} that we considered here, which lead to same decomposition \eref{Ein}.

\section{Summary and Discussion}
\label{summary}

We have extended the ADM constraint propagation equations, using the Frittelli method \cite{Frit97}, to 
 generic metric $f(R)$ gravity represented as a BD theory. For direct comparison with published results, we wrote our 
general evolution equations of the constraints (defined via Equations~\eref{Hamiltonian_prim} and~\eref{Momentum_prim})
in the same form as the original equations given in \cite{Frit97}. This mathematical form, 
given by Equations~\eref{con_prop_ham} and \eref{con_prop_mom},
combines both spacetime and purely spatial objects. 
To make transparent the connection between these
 equations and the language of the 3+1 decomposition of spacetime, we
 cast Equations \eref{con_prop_ham} and \eref{con_prop_mom} in pure 3+1 form, i.e., in a form that involves only scalar 
and purely spatial objects and their derivatives (see Equations~\eref{H_evol_final} and~\eref{Mlowj_evol_final}). 
The 3+1 form is the mathematical
 form the evolution equations of the constraints would take on, if one had employed a ``brute force'' 3+1 approach 
for performing this derivation. The brute force approach requires prior knowledge of the exact 3+1 equations and is much more 
involved.  

 The main result of this work is that the mathematical form of the constraint
 propagation equations is the same for both vacuum and non-vacuum spacetimes, as well as for any (viable) form of the function $f(R)$, 
provided that $T_{\mu\nu}^{(m)}$ and $T_{\mu\nu}^{(\phi)}$ (see \Sref{3p1_intro}) are absorbed properly in the definition of the constraints. 
We have also argued that the mathematical form of the evolution of the constraints for 3+1 $f(R)$ gravity
 in the Jordan frame remains the same as that of the 
BD-equivalent 3+1 $f(R)$ gravity. This result holds true in the Einstein frame, too, if the spacetime foliation is chosen based on 
the Einstein metric $\tilde g_{\mu\nu}$ and not the physical (Jordan) metric $g_{\mu\nu}$. 
Finally, a comparison between our equations and previous GR results, using the 3+1 approach,
 shows that all expressions for the constraint propagation equations agree.

We end this work by pointing out that the 3+1 BD-equivalent $f(R)$ equations can be incorporated in current numerical relativity codes 
with only minor additional effort. For example, the minimum requirement for studying vacuum spacetimes in viable $f(R)$ models 
is to include the contribution of the scalar field stress-energy tensor $T_{\mu\nu}^{(\phi)}$ (see \Sref{3p1_intro}) 
and implement a scalar field solver in the form of Equations \eref{Pievol}-\eref{Qievol}. As in GR, it is almost certain that the stability 
of numerical implementations of the fully non-linear equations of $f(R)$ gravity will be sensitive to the formulation used.  
Given that the structure of the $f(R)$ constraint propagation equations is fundamentally the same as that of the ADM formulation, 
we believe that dynamical $f(R)$ simulations will benefit from formulations such as the Baumgarte-Shapiro-Shibata-Nakamura (BSSN;
\cite{ShibNakamBSSN,BaumShapirBSSN}) approach or the generalized harmonic decomposition \cite{Pretorius2005a}. If these formalisms fail, 
other approaches such as those proposed in \cite{PADM,Calabrese,Fiske} may prove useful. We hope that this work will serve as a starting point
for relativists to develop fully dynamical codes for viable $f(R)$ models.

\ack

We would like to thank Carlos Cunha for useful conversations. 
This
paper was supported in part by NSF Grants PHY06-50377 and
 PHY09-63136 as well as NASA Grants
NNX07AG96G and NNX10AI73G to the University of
Illinois at Urbana-Champaign. Ignacy Sawicki acknowledges support by 
the DFG through TRR33 ``The Dark Universe".

\section*{References}
\bibliographystyle{unsrt} 
\bibliography{notes}

\end{document}